\documentclass[aps,prc,superscriptaddress,showpacs,nofootinbib,floatfix,twocolumn]{revtex4}

\usepackage{graphicx}   

\def\pt{$p_T$}
\def\pts{$p_T$ }

\def\gevc{~GeV/$c$}
\def\gevcs{~GeV/$c$ }
\def\raa{$R_{\rm AA}$}
\def\raas{$R_{\rm AA}$ }

\def\pp{$p$+$p$ }
\def\piz{$\pi^{0}$}
\def\pizs{$\pi^{0}$ }
\def\h{$\eta$}
\def\hs{$\eta$ }

\def\mean#1{\left<#1\right>}

\begin{document}

\title{Transverse momentum dependence of $\eta$ meson suppression \\
       in Au+Au collisions at $\sqrt{s_{NN}}=200$ GeV
       }

\newcommand{\abilene}{Abilene Christian University, Abilene, Texas 79699, USA}
\newcommand{\acadsin}{Institute of Physics, Academia Sinica, Taipei 11529, Taiwan}
\newcommand{\banaras}{Department of Physics, Banaras Hindu University, Varanasi 221005, India}
\newcommand{\barc}{Bhabha Atomic Research Centre, Bombay 400 085, India}
\newcommand{\bnlcoll}{Collider-Accelerator Department, Brookhaven National Laboratory, Upton, New York 11973-5000, USA}
\newcommand{\bnlphys}{Physics Department, Brookhaven National Laboratory, Upton, New York 11973-5000, USA}
\newcommand{\caucr}{University of California - Riverside, Riverside, California 92521, USA}
\newcommand{\charlesczech}{Charles University, Ovocn\'{y} trh 5, Praha 1, 116 36, Prague, Czech Republic}
\newcommand{\chonbuk}{Chonbuk National University, Jeonju 561-756, Korea}
\newcommand{\ciae}{China Institute of Atomic Energy (CIAE), Beijing, People's Republic of China}
\newcommand{\cns}{Center for Nuclear Study, Graduate School of Science, University of Tokyo, 7-3-1 Hongo, Bunkyo, Tokyo 113-0033, Japan}
\newcommand{\colorado}{University of Colorado, Boulder, Colorado 80309, USA}
\newcommand{\columbia}{Columbia University, New York, New York 10027 and Nevis Laboratories, Irvington, New York 10533, USA}
\newcommand{\czechtech}{Czech Technical University, Zikova 4, 166 36 Prague 6, Czech Republic}
\newcommand{\dapnia}{Dapnia, CEA Saclay, F-91191, Gif-sur-Yvette, France}
\newcommand{\debrecen}{Debrecen University, H-4010 Debrecen, Egyetem t{\'e}r 1, Hungary}
\newcommand{\elte}{ELTE, E{\"o}tv{\"o}s Lor{\'a}nd University, H - 1117 Budapest, P{\'a}zm{\'a}ny P. s. 1/A, Hungary}
\newcommand{\ewha}{Ewha Womans University, Seoul 120-750, Korea}
\newcommand{\fit}{Florida Institute of Technology, Melbourne, Florida 32901, USA}
\newcommand{\fsu}{Florida State University, Tallahassee, Florida 32306, USA}
\newcommand{\gsu}{Georgia State University, Atlanta, Georgia 30303, USA}
\newcommand{\hiroshima}{Hiroshima University, Kagamiyama, Higashi-Hiroshima 739-8526, Japan}
\newcommand{\ihepprot}{IHEP Protvino, State Research Center of Russian Federation, Institute for High Energy Physics, Protvino, 142281, Russia}
\newcommand{\illuiuc}{University of Illinois at Urbana-Champaign, Urbana, Illinois 61801, USA}
\newcommand{\instpasczech}{Institute of Physics, Academy of Sciences of the Czech Republic, Na Slovance 2, 182 21 Prague 8, Czech Republic}
\newcommand{\isu}{Iowa State University, Ames, Iowa 50011, USA}
\newcommand{\jinrdubna}{Joint Institute for Nuclear Research, 141980 Dubna, Moscow Region, Russia}
\newcommand{\jyvaskyla}{Helsinki Institute of Physics and University of Jyv{\"a}skyl{\"a}, P.O.Box 35, FI-40014 Jyv{\"a}skyl{\"a}, Finland}
\newcommand{\kek}{KEK, High Energy Accelerator Research Organization, Tsukuba, Ibaraki 305-0801, Japan}
\newcommand{\kfki}{KFKI Research Institute for Particle and Nuclear Physics of the Hungarian Academy of Sciences (MTA KFKI RMKI), H-1525 Budapest 114, POBox 49, Budapest, Hungary}
\newcommand{\korea}{Korea University, Seoul 136-701, Korea}
\newcommand{\kurchatov}{Russian Research Center ``Kurchatov Institute", Moscow, Russia}
\newcommand{\kyoto}{Kyoto University, Kyoto 606-8502, Japan}
\newcommand{\labllr}{Laboratoire Leprince-Ringuet, Ecole Polytechnique, CNRS-IN2P3, Route de Saclay, F-91128, Palaiseau, France}
\newcommand{\lawllnl}{Lawrence Livermore National Laboratory, Livermore, California 94550, USA}
\newcommand{\losalamos}{Los Alamos National Laboratory, Los Alamos, New Mexico 87545, USA}
\newcommand{\lpc}{LPC, Universit{\'e} Blaise Pascal, CNRS-IN2P3, Clermont-Fd, 63177 Aubiere Cedex, France}
\newcommand{\lund}{Department of Physics, Lund University, Box 118, SE-221 00 Lund, Sweden}
\newcommand{\maryland}{University of Maryland, College Park, Maryland 20742, USA}
\newcommand{\mass}{Department of Physics, University of Massachusetts, Amherst, Massachusetts 01003-9337, USA }
\newcommand{\muenster}{Institut fur Kernphysik, University of Muenster, D-48149 Muenster, Germany}
\newcommand{\muhlenberg}{Muhlenberg College, Allentown, Pennsylvania 18104-5586, USA}
\newcommand{\myongji}{Myongji University, Yongin, Kyonggido 449-728, Korea}
\newcommand{\nagasaki}{Nagasaki Institute of Applied Science, Nagasaki-shi, Nagasaki 851-0193, Japan}
\newcommand{\newmex}{University of New Mexico, Albuquerque, New Mexico 87131, USA }
\newcommand{\nmsu}{New Mexico State University, Las Cruces, New Mexico 88003, USA}
\newcommand{\ornl}{Oak Ridge National Laboratory, Oak Ridge, Tennessee 37831, USA}
\newcommand{\orsay}{IPN-Orsay, Universite Paris Sud, CNRS-IN2P3, BP1, F-91406, Orsay, France}
\newcommand{\peking}{Peking University, Beijing, People's Republic of China}
\newcommand{\pnpi}{PNPI, Petersburg Nuclear Physics Institute, Gatchina, Leningrad region, 188300, Russia}
\newcommand{\riken}{RIKEN Nishina Center for Accelerator-Based Science, Wako, Saitama 351-0198, JAPAN}
\newcommand{\rikjrbrc}{RIKEN BNL Research Center, Brookhaven National Laboratory, Upton, New York 11973-5000, USA}
\newcommand{\rikkyo}{Physics Department, Rikkyo University, 3-34-1 Nishi-Ikebukuro, Toshima, Tokyo 171-8501, Japan}
\newcommand{\saispbstu}{Saint Petersburg State Polytechnic University, St. Petersburg, Russia}
\newcommand{\saopaulo}{Universidade de S{\~a}o Paulo, Instituto de F\'{\i}sica, Caixa Postal 66318, S{\~a}o Paulo CEP05315-970, Brazil}
\newcommand{\seoulnat}{Seoul National University, Seoul 151-742, Korea}
\newcommand{\stonybrkc}{Chemistry Department, Stony Brook University, Stony Brook, SUNY, New York 11794-3400, USA}
\newcommand{\stonycrkp}{Department of Physics and Astronomy, Stony Brook University, SUNY, Stony Brook, New York 11794, USA}
\newcommand{\subatech}{SUBATECH (Ecole des Mines de Nantes, CNRS-IN2P3, Universit{\'e} de Nantes) BP 20722 - 44307, Nantes, France}
\newcommand{\tenn}{University of Tennessee, Knoxville, Tennessee 37996, USA}
\newcommand{\titech}{Department of Physics, Tokyo Institute of Technology, Oh-okayama, Meguro, Tokyo 152-8551, Japan}
\newcommand{\tsukuba}{Institute of Physics, University of Tsukuba, Tsukuba, Ibaraki 305, Japan}
\newcommand{\vandy}{Vanderbilt University, Nashville, Tennessee 37235, USA}
\newcommand{\waseda}{Waseda University, Advanced Research Institute for Science and Engineering, 17 Kikui-cho, Shinjuku-ku, Tokyo 162-0044, Japan}
\newcommand{\weizmann}{Weizmann Institute, Rehovot 76100, Israel}
\newcommand{\yonsei}{Yonsei University, IPAP, Seoul 120-749, Korea}
\affiliation{\abilene}
\affiliation{\acadsin}
\affiliation{\banaras}
\affiliation{\barc}
\affiliation{\bnlcoll}
\affiliation{\bnlphys}
\affiliation{\caucr}
\affiliation{\charlesczech}
\affiliation{\chonbuk}
\affiliation{\ciae}
\affiliation{\cns}
\affiliation{\colorado}
\affiliation{\columbia}
\affiliation{\czechtech}
\affiliation{\dapnia}
\affiliation{\debrecen}
\affiliation{\elte}
\affiliation{\ewha}
\affiliation{\fit}
\affiliation{\fsu}
\affiliation{\gsu}
\affiliation{\hiroshima}
\affiliation{\ihepprot}
\affiliation{\illuiuc}
\affiliation{\instpasczech}
\affiliation{\isu}
\affiliation{\jinrdubna}
\affiliation{\jyvaskyla}
\affiliation{\kek}
\affiliation{\kfki}
\affiliation{\korea}
\affiliation{\kurchatov}
\affiliation{\kyoto}
\affiliation{\labllr}
\affiliation{\lawllnl}
\affiliation{\losalamos}
\affiliation{\lpc}
\affiliation{\lund}
\affiliation{\maryland}
\affiliation{\mass}
\affiliation{\muenster}
\affiliation{\muhlenberg}
\affiliation{\myongji}
\affiliation{\nagasaki}
\affiliation{\newmex}
\affiliation{\nmsu}
\affiliation{\ornl}
\affiliation{\orsay}
\affiliation{\peking}
\affiliation{\pnpi}
\affiliation{\riken}
\affiliation{\rikjrbrc}
\affiliation{\rikkyo}
\affiliation{\saispbstu}
\affiliation{\saopaulo}
\affiliation{\seoulnat}
\affiliation{\stonybrkc}
\affiliation{\stonycrkp}
\affiliation{\subatech}
\affiliation{\tenn}
\affiliation{\titech}
\affiliation{\tsukuba}
\affiliation{\vandy}
\affiliation{\waseda}
\affiliation{\weizmann}
\affiliation{\yonsei}
\author{A.~Adare} \affiliation{\colorado}
\author{S.~Afanasiev} \affiliation{\jinrdubna}
\author{C.~Aidala} \affiliation{\mass}
\author{N.N.~Ajitanand} \affiliation{\stonybrkc}
\author{Y.~Akiba} \affiliation{\riken} \affiliation{\rikjrbrc}
\author{H.~Al-Bataineh} \affiliation{\nmsu}
\author{J.~Alexander} \affiliation{\stonybrkc}
\author{K.~Aoki} \affiliation{\kyoto} \affiliation{\riken}
\author{L.~Aphecetche} \affiliation{\subatech}
\author{Y.~Aramaki} \affiliation{\cns}
% ...
\collaboration{PHENIX Collaboration} \noaffiliation

\date{\today}

\begin{abstract}

New measurements by the PHENIX experiment at RHIC for $\eta$ 
production at midrapidity as a function of transverse momentum 
($p_T$) and collision centrality in $\sqrt{s_{NN}}=200$ GeV Au+Au 
and $p+p$ collisions are presented.  They indicate nuclear 
modification factors ($R_{\rm AA}$) which are similar both in 
magnitude and trend to those found in earlier $\pi^{0}$ 
measurements.  Linear fits to $R_{\rm AA}$ as a function of $p_T$ in 
5--20 GeV/$c$ show that the slope is consistent with zero within 
two standard deviations at all centralities although a slow rise 
cannot be excluded.  Having different statistical and systematic 
uncertainties, the $\pi^{0}$ and $\eta$ measurements are 
complementary at high $p_T$; thus, along with the extended $p_T$ 
range of these data they can provide additional constraints for 
theoretical modeling and the extraction of transport properties.

\end{abstract}

\pacs{25.75.Dw, 13.85.Qk, 13.20.Fc, 13.20.He}

\maketitle



Suppression of high \pts hadron production in Au+Au collisions at 
RHIC~\cite{ppg003,ppg080} and its absence in $d$+Au 
collisions~\cite{ppg028} provided the first direct evidence that an 
extremely dense medium is formed in heavy ion collisions at RHIC 
energies.  This suppression relative to the yield expected from the 
convolution of independent nucleon-nucleon scatterings, measured by 
the nuclear modification factor \raa, is now confirmed up to ~20\gevcs 
with identified \pizs and attributed to the energy loss of the hard 
scattered partons in the dense medium.  Several models with very 
different assumptions describe the magnitude of the observed \pizs 
suppression, but predict slightly different evolution with increasing 
\pt.  Calculations based on perturbative QCD (pQCD) and static plasma 
predict that the fractional parton energy loss decreases with \pts 
like $log(p_T)/p_T$ leading to a slow rise of the \raas with \pts (for 
a recent review see~\cite{bass2009}).  In contrast, some AdS/CFT 
calculations find that the fractional energy loss is proportional to 
\pt.  Therefore, \raas {\it decreases} with increasing transverse 
momentum~\cite{Liu2006,Horowitz2008,Chesler2009,Gubser2008}.  The 
universal upper bound model~\cite{Kharzeev2008} predicts that \raas 
remains almost independent of the energy of the original gluon or 
quark.  Other effects (modified nuclear parton distribution functions, 
Cronin-effect, modified fragmentation functions, the quark/gluon 
ratio) at given $x_T$ ($2p_T/\sqrt{s}$) can also 
dependence of \raa, and it is clear change the \pts dependence of 
\raa, and it is clear that a precise measurement of the evolution of 
\raas with \pts would help in confirming or rejecting classes of 
theories and putting tight constraints on the free parameters of the 
remaining ones.  The first rigorous attempt to confront the observed 
\pizs suppression with various pQCD-based parton energy loss 
calculations and to put quantitative constraints on the transport 
properties of the medium was made in~\cite{ppg079} using PHENIX \pizs 
data.  One intriguing result was that a linear fit with a slope 
consistent with zero described the evolution of \raas with \pts 
slightly better than any of the pQCD models predicting a slow rise.  
However, the large statistical and systematic uncertainties of the 
high \pts \pizs points prevented a clear distinction between constant 
or slowly rising \raa.

\begin{figure}[tbh]
\includegraphics[width=1.0\linewidth]{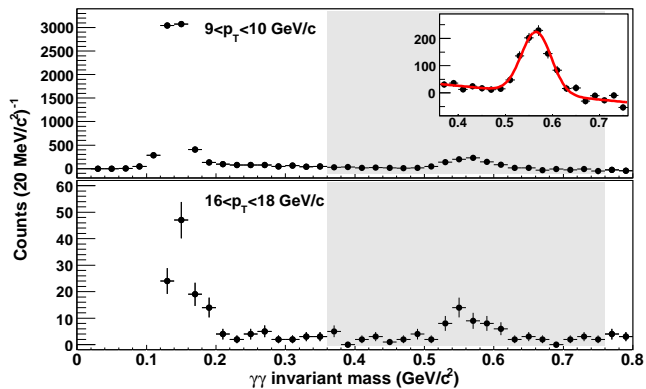}
\caption{\label{fig:fig1_invmass} (Color online) 
  $\gamma\gamma$ invariant mass distribution for two different pair 
  \pts bins  (minimum bias data).
  Top: 9$<$\pt$<$10 \gevc, the combinatorial background has been subtracted
  by using mixed events.  Note the large difference between \pizs and
  \hs raw yields.  Insert: the \hs region magnified.  
  Bottom: 16$<$\pt$<$18 \gevcs region, where mixed event subtraction
  is no longer necessary.  Also, here a cut on the $\gamma$-pair energy
  asymmetry, $\alpha<0.6$ has been applied, which greatly improves the
  signal/background ratio at the
  \hs peak but cuts into the lower part of the \pizs peak due to
  cluster merging.  
}
\end{figure}

One reason the \pizs data~\cite{ppg080} allow such ambiguous 
interpretations is that the experimental uncertainties rise rapidly as 
we move to higher \pts ($>$12--14\gevc), due to ``shower merging,'' as 
explained below.  In the case of the \hs this problem is absent for 
\pts up to 50\gevc, significantly beyond the \pts range expected to be 
accessible at RHIC.  While the yield of the actually reconstructed \hs 
mesons is smaller except at the highest \pt, the improvement in 
systematic uncertainties can help provide better constraints in 
comparisons to theory at high \pts and thus complement the \pizs 
results.  Of course some caution in interpreting the results is 
warranted: while both \pizs and \hs consist of light quarks, \hs does 
have a hidden strangeness ($s\bar{s}$) content so it is not {\it a 
priori} obvious that the \pizs and \hs results are interchangeable.  
Earlier measurements~\cite{ppg055} have shown that at least up to 
12\gevc, the \pizs and \hs nuclear modification factors in Au+Au agree 
within uncertainties and the $\eta/\pi^0$ ratio is constant for 
\pt$\geq$4 \gevcs in \pp~\cite{ppg055}.  Using recent, more precise 
measurements in PHENIX, we will re-examine whether \pizs and \hs 
production at midrapidity is indeed similar and study the asymptotic 
behavior of \raa.


This analysis used 3.25B minimum bias (MB) $\sqrt{s_{NN}}=200$ GeV 
Au+Au events, corresponding to 0.511 nb$^{-1}$ recorded in 2007 as 
well as 429M minimum bias (18.7 nb$^{-1}$) and 2.06B triggered (6.90 
pb$^{-1}$) $\sqrt{s}=200$ GeV \pp events recorded in 2006 in the 
PHENIX experiment at RHIC.  Both the Au+Au and \pp data sets were 
analyzed using the same analysis chain and cuts; thus, some of the 
systematic uncertainties cancel when we calculate the nuclear 
modification factor \raas for Au+Au.

Collision centrality in Au+Au has been established by the beam-beam 
counters~\cite{nimbbc} (BBC, $3.0<|\eta|<3.9$).  A Glauber-model Monte 
Carlo along with a simulation of the BBC response was used to estimate 
the average number of participating nucleons ($N_{\rm part}$) and 
binary nucleon-nucleon collisions ($N_{\rm coll}$) for each centrality 
bin~\cite{ppg026}.

\begin{table}[tbh]
\caption{\label{table:syst_spec_raa}
    Typical systematic uncertainties on \hs spectra and \raa.  See
 text for explanation of error types.
 }
\begin{ruledtabular}
\begin{tabular}{ccccc}
  Source                 & Type &  Au+Au & \pp & \raas \\ \hline
  raw yield              &  B   &  7\%    & 3\% & 6.3\% \\
  acceptance variations  &  B   &  1.5\%  & 1.5\% & 2.1\% \\
  photon PID             &  B   &  3\%    & 3\% & 3\% \\
  acceptance$\times$efficiency  &  A   &  3\%    & 3\% & 4.2\% \\
  energy scale           &  B   &  8\%    & 8\% & 11.3\% \\
  conversion (HBD)       &  C   &  1.3\%  & N/A & 1.3\% \\
  conversion (other)     &  C   &  5\%    & 5\% & N/A \\
  BBC cross section      &  C   &  N/A    & 9.7\% & 9.7\% \\
  BBC efficiency         &  C   &  N/A    & 3.8\% & 3.8\% \\
  ERT norm.              &  C   &  N/A    & 6.2\% & 6.2\% \\
 \end{tabular}
\end{ruledtabular}
\end{table}

The \hs mesons were measured via their $\eta\rightarrow\gamma\gamma$ 
decay channel.  The photons were reconstructed in the 
lead-scintillator (PbSc) sectors of the PHENIX Electromagnetic 
Calorimeter (EMCal)~\cite{nimemc} covering 3/8 of the full azimuth and 
$-0.35<\eta<0.35$ in pseudorapidity, and the \hs yield was extracted 
from two-photon invariant mass distributions.  This analysis is 
similar to the one described in~\cite{ppg055,ppg054}.  There are three 
important differences.  In the case of \pizs starting around 
\pt=12\gevcs the minimum opening angle of the two decay photons is 
small enough for the photon showers to merge and become 
indistinguishable.  As \pts increases, this effect leads to an 
increasing loss of observed \piz, resulting in large corrections and 
corresponding systematic uncertainties (which are in fact the dominant 
systematic uncertainties at high \pt).  Since the mass of the \hs is 
about four times larger than the \piz, this is not a problem for the 
\hs measurement up to \pt$\sim$50 \gevc.  On the other hand the 
observable \hs rates are much lower at low and medium \pt, as seen in 
the invariant mass distributions in Fig.~\ref{fig:fig1_invmass}, 
because of the smaller branching ratio into two photons (39\%) and the 
small \h/\piz$\approx$0.5 production ratio.  The raw yields become 
comparable only around 20 \gevc.  Finally, in the \hs analysis we 
applied an $\alpha<0.6$ photon pair energy asymmetry cut (as opposed 
to $\alpha<0.8$ for \piz) in order to improve the signal/background 
ratio in the \hs region.

\begin{figure}[tb]
\includegraphics[width=1.0\linewidth]{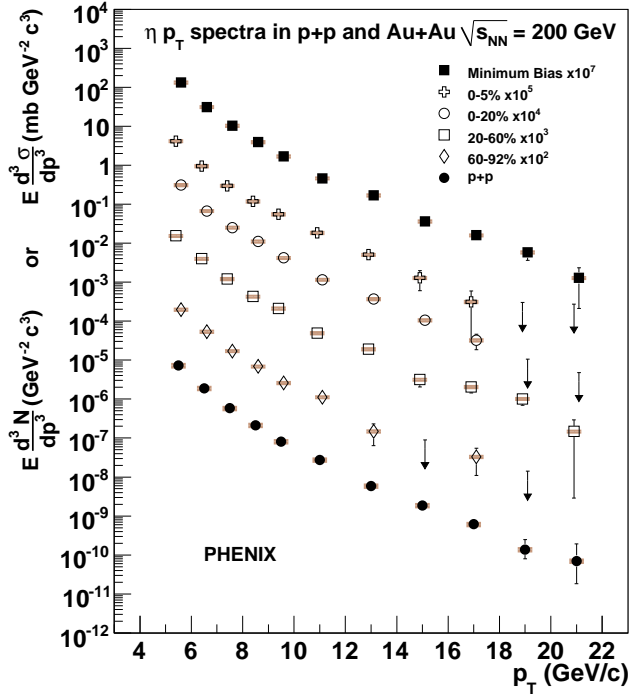}
\caption{\label{fig:fig2_etafinalspec} (Color online) Cross 
section of $p+p\rightarrow\eta+X$ from the 2006 \pp data set (solid
circles) and \hs invariant yield in Au+Au collisions of various
centralities (open symbols) and minimum bias (solid squares) 
from the 2007 data set.  \pp is shown at the true \pts value, all
other spectra are shifted alternately by $\pm$0.1\gevcs for better
visibility of the error bars and upper limits.}
\end{figure}

\begin{table}[hb]
\caption{Parameters of the power-law fits $A/p_T^n$ 
for Au+Au and \pp.  \
The errors used for fit are the statistical and \pt-uncorrelated
(Type A) systematic uncertainties added in quadrature.  The \pts
range of the fits is 5--22 \gevc.}
\label{table:specfitparams}
\begin{ruledtabular}
\begin{tabular}{cccc}
System/Cent. & $A$ & $n$ & $\chi^2$/NDF \\
\hline
Au+Au 0--5\,\% & 27.2$\pm$11.9 & 7.90$\pm$0.22 & 3.1/7 \\
Au+Au 0--10\,\% & 17.6$\pm$5.5 & 7.77$\pm$0.15 & 10.6/8 \\
Au+Au 10--20\,\% & 19.1$\pm$5.9 & 7.89$\pm$0.16 & 10.2/9 \\
Au+Au 0--20\,\% & 18.5$\pm$4.3 & 7.84$\pm$0.12 & 10.5/7 \\
Au+Au 20--40\,\% & 17.3$\pm$4.2 & 8.01$\pm$0.12 & 17.2/8 \\
Au+Au 40--60\,\% & 9.53$\pm$2.65 & 8.05$\pm$0.15 & 5.5/8 \\
Au+Au 20--60\,\% & 14.5$\pm$2.5 & 8.07$\pm$0.08 & 11.2/9 \\
Au+Au 60--92\,\% & 1.13$\pm$0.40 & 7.78$\pm$0.18 & 2.98/6 \\
Au+Au MinBias & 10.4$\pm$1.4 & 8.04$\pm$0.08 & 9.41/9 \\
\pp & 8.84$\pm$0.99 & 8.21$\pm$0.05 & 8.33/9 \\
\end{tabular}
\end{ruledtabular}
\end{table}

The raw \hs yield is always counted by integrating the histogram bin 
content in the \hs mass window (typically $\pm$30 MeV/$c^2$), but the 
way we treat the underlying combinatorial background varies as a 
function of \pt.  In Au+Au up to 10 \gevc, mixed event subtraction is 
used.  The \hs region is then fitted with a polynomial and Gaussian 
(see insert in Fig.~\ref{fig:fig1_invmass}) to estimate the residual 
background.  When the signal/background ratio reaches 1.0, already in 
the 7--10 \gevcs range, depending on centrality, mixed event 
subtraction is no longer needed; a polynomial and Gaussian fit is used 
on the original invariant mass distribution to estimate the 
background.  At even higher \pts (12--16 \gevc) we estimate the 
residual background under the peak simply from the average bin content 
of the sidebands (the regions above and below the peak).

\begin{figure}[tbh]
\includegraphics[width=1.0\linewidth]{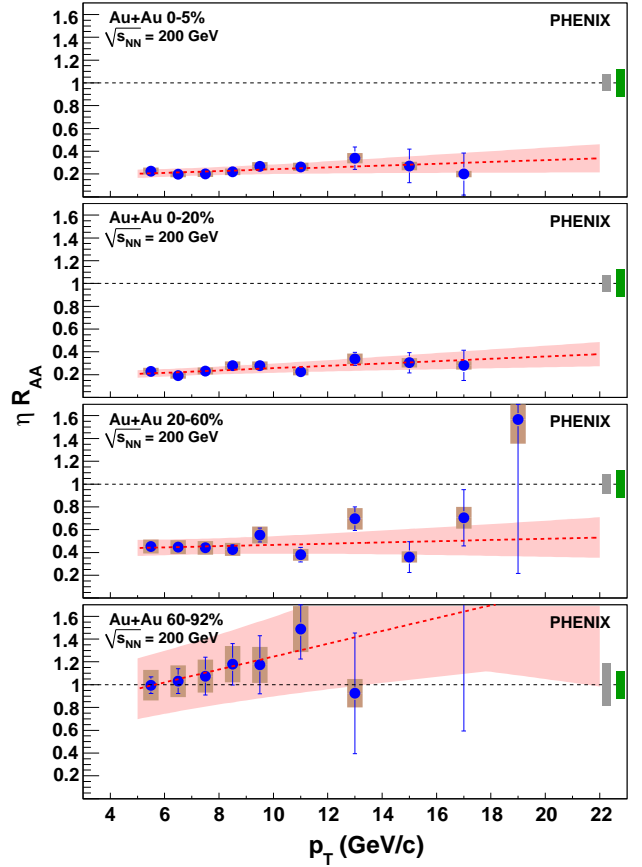}
\caption{\label{fig:fig3_etaraawithfit} (Color online) 
Nuclear modification factor for \hs at various centralities,
calculated using the measured \pp points.  Dark
(green) band around 1 indicates the absolute normalization error from
$p$+$p$, light (grey) band is the (centrality dependent) absolute
normalization error from Au+Au.  Error bars include statistical and
\pt-uncorrelated systematic errors.  Also shown: linear fits to the
data with 1$\sigma$ error bands.
}
\end{figure}

Systematic uncertainties are classified into three types: Type A is 
$p_T$-uncorrelated (``point-by-point'') and for the purposes of 
fitting and plotting, is added in quadrature to the statistical 
errors.  Type C is the overall normalization uncertainty allowing all 
points to move by the same fraction up or down.  Type B is all other 
\pt-correlated uncertainties (including the cases where the shape of 
the correlation function is not known).  
Table~\ref{table:syst_spec_raa} lists typical uncertainties on the 
spectra and \raa.  ``Conversion (HBD)'' stands for loss due to photon 
conversion in the Hadron Blind Detector, which was present in one of 
the two central arms during the 2007 (Au+Au) data taking.  ``ERT 
norm.'' stands for the normalization uncertainty of the EMCal-RICH 
Trigger, selecting high \pts photons and electrons.  ``Acceptance 
variations'' are small day-by-day changes of dead areas in the 
detector and thus are independent for the \pp and Au+Au runs.  The 
systematic uncertainties on raw yield, photon PID and conversion 
(other) are common in \pp and Au+Au, and hence were partially 
cancelled out in the $R_{\rm AA}$ calculation.

\begin{figure}[tbh]
\includegraphics[width=1.0\linewidth]{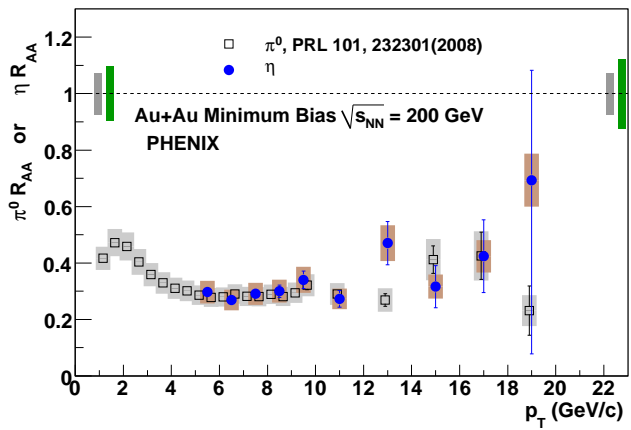}
\caption{\label{fig:fig4_etapi0raacomp} (Color online) 
Nuclear modification factor \raas for \pizs 
(open squares, points shifted for clarity, data from~\cite{ppg080}) 
and \hs (solid circles, this analysis) in MB Au+Au collisions.
Error bars include statistical and \pt-uncorrelated systematic errors,
bands show \pt-correlated systematic errors.  The pair of bands at
\raa=1 are the absolute normalization error for \pp (larger, dark) and
Au+Au (lighter) for \pizs (left) and \hs (right).
}
\end{figure}


Cross sections for \pp$\rightarrow\eta+X$ and invariant yield of 
inclusive \hs production in Au+Au collisions for different 
centralities are shown in Fig.~\ref{fig:fig2_etafinalspec}.  They 
cover the $5<p_T<22$\gevcs range and five orders of magnitude in cross 
section (invariant yield).  The overall normalization uncertainties 
(Type C) are 13\% for \pp and 5\% for Au+Au.  Parameters of simple 
power-law fits ($A/p_T^n$) to various, partially overlapping 
centrality selections, including ones not shown in 
Fig.~\ref{fig:fig2_etafinalspec}, are given in 
Table~\ref{table:specfitparams}.  Fits include all available points in 
the $5<$\pt$<22$\gevcs range but exclude upper limits.  Only 
statistical and \pt-uncorrelated uncertainties were used in the fits.  
Note that for \pizs in Au+Au collisions the power $n$ was consistent 
within uncertainties at all centralities~\cite{ppg080} ranging from 
8.00$\pm$0.12 in 0--5\% to 8.06$\pm$0.08 in 80--92\%, and for \pizs in 
\pp the power $n$ was 8.22$\pm$0.09.  In this measurement we find that 
for \hs production $p+p\rightarrow\eta+X$ the power $n$ is the same as 
it was for \piz.  The powers obtained for \hs in Au+Au are also 
consistent with those from \pizs within two standard deviations.

The nuclear modification factor \raas is defined as \[R_{\rm AA} = 
\frac{1/{N_{\rm evt}} dN/dydp_T}{\mean{T_{\rm AB}}d\sigma_{pp}/dydp_T} 
\] where $\sigma_{pp}$ is the production cross section of the particle 
in \pp collisions, and $\mean{T_{\rm AB}}$ is the nuclear thickness 
function averaged over a range of impact parameters for the given 
centrality, calculated within a Glauber model~\cite{Miller2007}.  When 
calculating \raa, the measured \pp points are used.  \raas for \hs 
production is shown in Fig.~\ref{fig:fig3_etaraawithfit} for four 
centralities, along with linear fits to \raa.  Fit parameters are 
listed in Table~\ref{table:raafitparams}.  In the measured \pts range 
we observe strong suppression in all but the most peripheral 
collisions.  As shown in Fig.~\ref{fig:fig4_etapi0raacomp}, for the 
minimum bias case the suppression is quite comparable to the one 
observed for \piz, and above 13\gevcs the (relative) systematic errors 
are smaller.

\begin{figure}[tbh]
\includegraphics[width=1.0\linewidth]{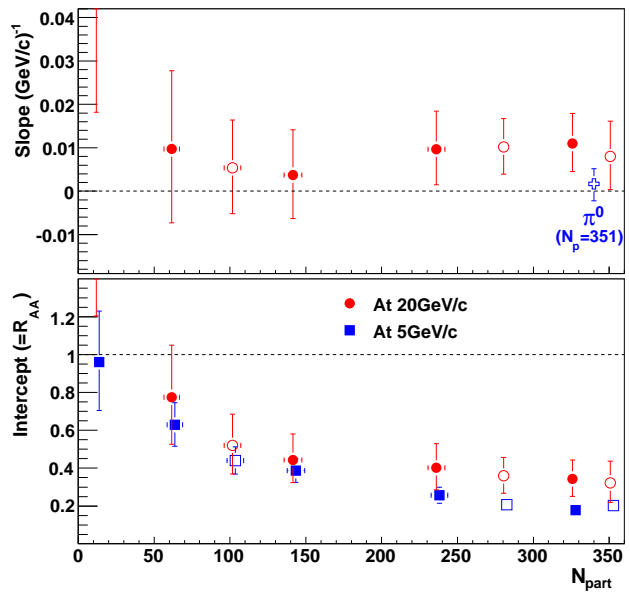}
\caption{\label{fig:fig5_raaslopeintercept} (Color online) 
Top: slopes of the linear fits 
(like the ones shown in Fig.~\ref{fig:fig3_etaraawithfit}) along
with the fitting errors.  Centrality is shown in terms of
participating nucleons $N_{\rm part}$.  Open symbols are overlapping,
solid symbols are non-overlapping centrality bins 
(0--10\%, 10--20\%, 20--40\%, 40--60\% and 60--92\%).
Also shown: slope of the linear fit to 0--5\% \pizs 
data~\cite{ppg079},
shifted for better visibility.
Bottom: value of \raas calculated from the fit at 5\gevcs (blue) 
and 20\gevcs (red).
}
\end{figure}

\begin{figure}[bh]
\includegraphics[width=1.0\linewidth]{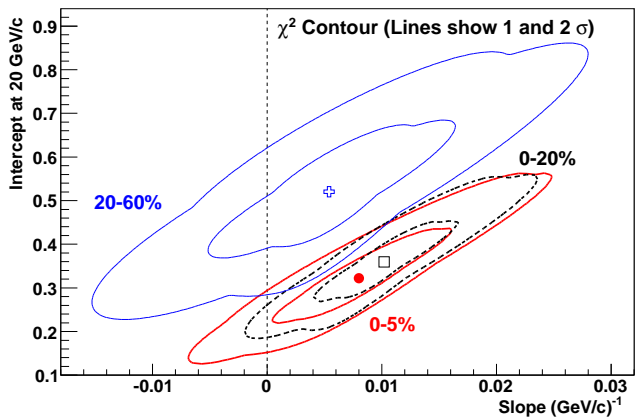}
\caption{\label{fig:fig6_slopecontour} (Color online) 
One and two standard deviation $\chi^2$ contours 
of the linear fits to \raas in Au+Au collisions 
for 0--5\%, 0--20\% and 20--60\% centralities.  }
\end{figure}

\begin{table}[htb]
\caption{Parameters from linear function fit to $\eta$ $R_{\rm AA}$.}
\label{table:raafitparams}
\begin{ruledtabular}
\begin{tabular}{cccc}
Centrality & $N_{\rm part}$ & Slope & $\chi^2$/NDF \\
\hline\hline
0--5\,\% & 351 & 0.008$\pm$0.008 & 2.77/7 \\
0--10\,\% & 326 &  0.011$\pm$0.007 & 9.79/7 \\
10--20\,\% & 236 &  0.010$^{+0.009}_{-0.008}$ & 11.7/8 \\
0--20\,\% & 280 &  0.010$^{+0.007}_{-0.006}$ & 10.8/7 \\
20--40\,\% & 142 &  0.004$\pm$0.010 & 15.7/8 \\
40--60\,\% & 61.6 &  0.010$^{+0.018}_{-0.017}$ & 4.64/7 \\
20--60\,\% & 102 &  0.005$\pm$0.011 & 11.7/8 \\
60--92\,\% & 11.8 &  0.056$^{+0.043}_{-0.038}$ & 1.52/6 \\
MinBias & 109 &  0.006$\pm$0.007 & 10.1/8 \\
\end{tabular}
\end{ruledtabular}
\end{table}

Based upon the most central (0--5\%) collisions in~\cite{ppg079} we 
found that the \pizs \raas is consistent with a completely flat \pts 
dependence when fitted in the $5<p_T<18$\gevcs region, namely the 
slope of a linear fit was $m=0.0017^{+0.0035}_{-0.0039}\,c$/GeV.  
Fitting the current \hs \raas data with straight lines gives the 
slopes and uncertainties listed in Table~\ref{table:raafitparams} and 
shown in Fig.~\ref{fig:fig5_raaslopeintercept} where centrality is 
expressed in terms of participating nucleons $N_{\rm part}$.  All 
slopes are consistent with zero; the largest deviation is less than 
$2\sigma$ (for the 0--20\% centrality bin).  One and two standard 
deviation $\chi^2$ contours for selected centrality bins are shown in 
Fig.~\ref{fig:fig6_slopecontour}.  For 0--5\% centrality we repeated 
the linear fits using only the first $3,4,...,(n-1)$ points and found 
that the slope already stabilizes around its final value with the 
first few points; data above 10 \gevcs improve the significance but 
barely change the central value itself.  The same is true for other 
centralities.

While the above result indicates that \raas for \hs is consistent with 
a \pt-independent, constant value, and disfavors a {\it decreasing} 
\raa, a slow rise ($\sim$0.01\,$c$/GeV) of $R_{\rm AA}$ with 
increasing \pts cannot be excluded.  In fact, a detailed statistical 
analysis, comparing to various theories like the study done for \pizs 
in~\cite{ppg079} is necessary once theoretical calculations of \hs 
production are available.  However, {\it assuming} the linear 
dependence we can calculate the \raas values at 5 \gevcs (where the 
suppression is already at its maximum) and 20 \gevc; the results are 
shown in the bottom panel of Fig.~\ref{fig:fig5_raaslopeintercept}.


In summary, we measured invariant yields of \hs in $\sqrt{s_{NN}}=200$ 
GeV Au+Au collisions at various centralities, as well as the \hs 
production cross section in $\sqrt{s}=200$ GeV \pp collisions in the 
$5<$\pt$<22$ \gevcs transverse momentum range using the PbSc 
calorimeter of the PHENIX experiment at RHIC.  The nuclear 
modification factor for \hs in minimum bias collisions is consistent 
with earlier \pizs results.  In conclusion, linear fits to \raas as a 
function of \pts indicate that \raas is consistent with constant at 
all centralities, although a slow rise cannot be excluded.



We thank the staff of the Collider-Accelerator and 
Physics Departments at BNL for their vital contributions.  
We acknowledge support from 
the Office of Nuclear Physics in DOE Office of Science, NSF,  
and a sponsored research grant from Renaissance Technologies (USA),
MEXT and JSPS (Japan), 
CNPq and FAPESP (Brazil), 
NSFC (China), 
MSMT (Czech Republic),
IN2P3/CNRS and CEA (France), 
BMBF, DAAD, and AvH (Germany), 
OTKA (Hungary), 
DAE and DST (India), 
ISF (Israel), 
NRF (Korea), 
MES, RAS, and FAAE (Russia),
VR and KAW (Sweden), 
U.S. CRDF for the FSU, 
US-Hungary Fulbright, 
and US-Israel BSF.


\begin{references}

\bibitem{ppg003} 
  K. Adcox {\it et al.},
  Phys. Rev. Lett. {\bf 88}, 022301 (2001).

\bibitem{ppg080} 
  A. Adare {\it et al.},
  Phys. Rev. Lett. {\bf 101}, 232301 (2008).

\bibitem{ppg028} 
  S. S. Adler {\it et al.},
  Phys. Rev. Lett. {\bf 91}, 072303 (2003).

\bibitem{bass2009} 
  S.A. Bass, C. Gale, A. Majumder, C. Nonaka, G.-Y. Qin, T. Renk,
  J. Ruppert,
  Phys. Rev. C {\bf 79}, 024901 (2009).

\bibitem{Liu2006} 
  H. Liu, K. Rajagopal, U.A. Wiedemann,
  Phys. Rev. Lett. {\bf 97}, 182301 (2006).

\bibitem{Horowitz2008} 
  W.A. Horowitz, M. Gyulassy,
  Phys. Lett. B {\bf 666}, 320 (2008).

\bibitem{Chesler2009} 
  P.M. Chesler, K. Jensen, A. Karch, L.G. Yaffe,
  Phys. Rev. D {\bf 79}, 125015 (2009).

\bibitem{Gubser2008} 
  S.S. Gubser, D.R. Gulotta, S.S. Pufu, F.D. Rocha,
  Journ. High Energy Phys. 10, 052 (2008)

\bibitem{Kharzeev2008} 
  D.E. Kharzeev, arXiv:0806.0358 [nucl-th] (2008).

\bibitem{ppg079} 
  A. Adare {\it et al.}, 
  Phys. Rev. C. {\bf 77}, 064907 (2008).

\bibitem{ppg055} 
  S. S. Adler {\it et al.},
  Phys. Rev. C {\bf 75}, 024909 (2007).

\bibitem{ppg026} 
A. D. Martin, R. G. Roberts, W. J. Stirling, R. S. Thorne,
  Phys. Rev. C {\bf 69}, 034909 (2004).

\bibitem{nimbbc} M. Allen {\it et al.}, 
  Nucl. Instrum. Meth. {\bf A499}, 549 (2003).

\bibitem{nimemc} L. Aphecetche {\it et al.}, 
  Nucl. Instrum. Meth. {\bf A499}, 521 (2003).

\bibitem{ppg054} 
  S. S. Adler {\it et al.},
  Phys. Rev. C {\bf 76}, 034904 (2007).

\bibitem{Miller2007} 
  M. L. Miller, K. Reygers, S. J. Sanders and P. Steinberg,
  Annu. Rev. Nucl. Part. Sci. {\bf 57}, 205 (2007).


\end{references}


\end{document}